\documentclass{article}

\usepackage{amsmath,amsthm,upref}
\usepackage{amssymb,amsbsy}
\usepackage{amsfonts}

\begin{document}

\baselineskip 18pt
\begin{center} {\bf \large Generalizations of Nonlinear and Supersymmetric Classical Electrodynamics}
\bigskip
\bigskip
\\ \baselineskip 14pt
Steven Duplij \\
Department of Physics and Technology,\\
V.N. Karazin Kharkov National University,\\
Svoboda Sq. 4,\\
Kharkov 61077, Ukraine\\
{\it steven.a.duplij@univer.kharkov.ua\/}\\
\bigskip
Gerald A. Goldin \\
Departments
of Mathematics and Physics\\
Rutgers University, Busch Campus\\
Piscataway, NJ 08854, USA\\
{\it geraldgoldin@dimacs.rutgers.edu\/}\\
\bigskip
Vladimir M. Shtelen\\
Department of Mathematics\\
Rutgers University, Busch Campus\\
Piscataway, NJ 08854, USA\\
{\it shtelen@math.rutgers.edu\/}\\
{\ }\\
\end{center}

\begin{quote}\baselineskip 14pt
\noindent{\bf Abstract} We first write down a very general description of nonlinear
classical electrodynamics, making use of generalized
constitutive equations and constitutive tensors. Our approach
includes non-Lagrangian as well as Lagrangian theories, allows for
electromagnetic fields in the widest possible variety of media
(anisotropic, piroelectric, chiral and ferromagnetic), and
accommodates the incorporation of nonlocal effects. We
formulate electric-magnetic duality in terms of the
constitutive tensors. We then propose a supersymmetric
version of the general constitutive equations, in a superfield
approach.
\end{quote}

\baselineskip 14pt

\section{Introduction}

Maxwell's equations for the vector fields
$\mathbf{D}$ (electric displacement), $\mathbf{E}$
(electric field), $\mathbf{B}$ (magnetic induction), and $\mathbf{H}$
(magnetic field) in the vacuum or in media are well known to be incomplete.
The system is completed with {\it constitutive equations}, which
establish functional relations among these vector fields
\cite{lan/lif2,jackson}. Then $\mathbf{D}$ and $\mathbf{H}$ may be regarded as
constructs used to describe (via the constitutive equations) how the directly
observable fields $\mathbf{E}$ and $\mathbf{B}$ are produced by charge and
current densities. The explicit form of the constitutive equations depends
on the physical properties one assumes for the vacuum or for the
medium in which the fields occur; in particular, they are constrained
by the symmetry of the medium.

It has also long been known \cite{leb/lev},
though not always widely appreciated, that Maxwell's equations alone
(without constitutive equations) are consistent with both Lorentz symmetry
(for any value of the light speed $c$) and Galilei
symmetry. Thus one can specify a particular Lorentz- or Galilei-covariant
theory through the choice of constitutive equations. Linear
constitutive equations, taken together with Maxwell's equations, are
inconsistent with Galilei-covariant electrodynamics,
while a certain class of nonlinear equations are compatible with it \cite{gol/sht1}.
These observations were subsequently generalized to a class of
equations for nonabelian gauge fields, written using nonlinear
constitutive equations \cite{gol/sht2}. Interest in
nonlinearity and Maxwell's equations is heightened by
experimental results in nonlinear optics, such as
optical squeezing and slow light speed \cite{mara}, and by
theoretical ideas such as Born-Infeld theories of superstrings
\cite{bor/inf,frad,frad2,tsey}. It has recently been suggested
that a modified Born-Infeld Lagrangian,
proposed originally for the purpose of introducing a Galilean limit
in nonlinear electromagnetism, provides a way to
introduce a null string (i.e., zero tension) limit in a relativistic
theory of four-dimensional superstrings \cite{golmavsza}. We
also remark on the recent use of modifications of Maxwell's equations
as ``test theories'' in astrophysical observations, where upper
bounds to various sorts of possible deviations from known physical
laws can be established through measurement \cite{lamm1,lamm2}. Our approach
provides a very general framework for the construction of such
test theories, particularly in the direction of allowing for
dissipative effects. Our approach also accommodates various interesting limits
of physical constants.

This article continues the systematic exploration of nonlinear
 constitutive equations. We first write a very general form
 for such equations and the ``constitutive tensors'' that appear
 in them. As in earlier work, this  includes non-Lagrangian
 (i.e., dissipative) as well as Lagrangian theories, but it also allows for
the description of electromagnetic fields in the widest possible
variety of media, including anisotropic, piroelectric, chiral and
ferromagnetic media. Such a description also
accommodates the incorporation of nonlocal effects. We then
formulate electric-magnetic duality in terms of the
constitutive tensors. Finally, we propose a supersymmetric
version of the general constitutive equations, so as to obtain
a general, nonlinear supersymmetric electrodynamics within a
superfield approach. Some of our results were sketched briefly
in Ref. \cite{dupgolsht}; here we provide greater detail and
additional development.

\section{Maxwell's equations and nonlinear constitutive equations}

To set the context and specify notation, we review established
results in this section. Let us write Maxwell's equations in SI units,
\begin{align}
\operatorname{curl}\mathbf{E} &  =-\dfrac{\partial\mathbf{B}}{\partial
t},\ \ \ \ \ \ \operatorname{div}\mathbf{B}=0,\nonumber\\
\operatorname{curl}\mathbf{H} &  =\dfrac{\partial\mathbf{D}}{\partial
t}+\mathbf{j},\ \ \ \operatorname{div}\mathbf{D}=\rho,\label{max3a}%
\end{align}
where $\mathbf{j}$ and $\rho$ are current and charge densities. We
consider only flat spacetime, so that the metric is given by
the Minkowski tensor $\eta_{\mu\nu
}=\left(  1,-1,-1,-1\right)  $, $x^{\mu}=\left(  ct,x^{i}\right)  $,
$\mu,\nu, \dots \,=\,0,1,2,3$; $i,j, \dots \,=\,1,2,3$; with $\partial_{\mu}=\partial
\diagup\partial x^{\mu}\,=\,\left[  c^{-1}\partial\diagup\partial t,\nabla\right]
$. The antisymmetric Levi-Civita tensor is written $\varepsilon^{\mu
\nu\rho\sigma}$, with $\varepsilon^{0123}=1$. We denote six Lorentz invariants
constructed from $\mathbf{E},\,\mathbf{B},\,\mathbf{D},\,\mathbf{H}$ (in terms
of which other invariants may be written) by
\begin{align}
C_{1}  & =\mathbf{B}^{2}-\dfrac{1}{c^{2}}\mathbf{E}^{2},\ C_{2}=\mathbf{B}%
\cdot\mathbf{E,\ C}_{3}=\mathbf{D}^{2}-\dfrac{1}{c^{2}}\mathbf{H}^{2}%
,\ C_{4}=\mathbf{D}\cdot\mathbf{H},\nonumber\\
C_{5}  & =\mathbf{B}\cdot\mathbf{H}-\mathbf{E}\cdot\mathbf{D},\ C_{6}%
=\mathbf{B\cdot D}+\dfrac{1}{c^{2}}\mathbf{E\cdot H}.\label{cc}%
\end{align}
The constitutive equations relating $\mathbf{E}$, $\mathbf{B}$, $\mathbf{D}$ and
$\mathbf{H}$ can reduce the symmetry of Eqs. (\ref{max3a}) to the Lorentz or
Galilean groups; some other possibilities are considered in \cite{fus/sht/ser}.

In covariant notation we have the standard electromagnetic tensor fields%
\begin{equation}
F_{\mu\nu}=\left(\!\!
\begin{array}
[c]{cccc}%
0 & \frac{1}{c}\,E_{x} & \frac{1}{c}\,E_{y} & \frac{1}{c}\,E_{z}\\
-\frac{1}{c}\,E_{x} & 0 & -B_{z} & B_{y}\\
-\frac{1}{c}\,E_{y} & B_{z} & 0 & -B_{x}\\
-\frac{1}{c}\,E_{z} & -B_{y} & B_{x} & 0
\end{array}
\!\!\!\right),\,\, G_{\mu\nu}=\left(\!\!
\begin{array}
[c]{cccc}%
0 & cD_{x} & cD_{y} & cD_{z}\\
-cD_{x} & 0 & -H_{z} & H_{y}\\
-cD_{y} & H_{z} & 0 & -H_{x}\\
-cD_{z} & -H_{y} & H_{x} & 0
\end{array}
\!\!\!\right)\!;
\end{equation}
the Hodge dual tensors are $\tilde{F}^{\mu\nu}=\frac{1}{2}\varepsilon
^{\mu\nu\rho\sigma}F_{\rho\sigma}$ and $\tilde{G}^{\mu\nu}=\frac{1}
{2}\varepsilon^{\mu\nu\rho\sigma}G_{\rho\sigma}$; and Maxwell's
equations become
\begin{equation}
\partial_{\mu}\tilde{F}^{\mu\nu}=0,\ \ \ \ \ \ \partial_{\mu}G^{\mu\nu}%
=j^{\nu},\label{max4}%
\end{equation}
where $j^{\mu}=\left(  c\rho,\mathbf{j}\right)  $ is the 4-current.
The first equations in (\ref{max4}) imply $F_{\mu\nu}=\partial_{\mu
}A_{\nu}-\partial_{\nu}A_{\mu}$, where $A_{\mu}$ is an abelian gauge field;
but in general there is no similar representation for $G_{\mu\nu}$. The
field strength tensors $F_{\mu\nu}$ and $\tilde{F}^{\mu\nu}$ are physically
observable, in that their components can be inferred from measurement
of the force
$\,\mathbf{F}=q\left(  \mathbf{E}%
+\mathbf{v\times B}\right)$ on a test charge $q$ moving with velocity
$\mathbf{v}$. The tensors $G_{\mu\nu}$ and $\tilde{G}^{\mu\nu}$ originate
with the currents, and their relation to the observable fields is
determined by the properties of the medium (or the vacuum) within which
they are being described.

If we write constitutive equations for nonlinear electromagnetism in the form
$\mathbf{D}=\mathbf{D}\left(  \mathbf{E},\mathbf{B}\right),\,
\mathbf{H}=\mathbf{H}\left(  \mathbf{E},\mathbf{B}\right)$,
then for a Lorentz covariant theory, they must take the form \cite{fus/sht/ser}
\begin{align}
\mathbf{D} &  =M\left(  C_{1},C_{2}\right)  \mathbf{B}+\dfrac{1}{c^{2}%
}N\left(  C_{1},C_{2}\right)  \mathbf{E},\nonumber\\
\mathbf{H} &  =N\left(  C_{1},C_{2}\right)  \mathbf{B}-M\left(  C_{1}%
,C_{2}\right)  \mathbf{E},\label{con1}%
\end{align}
where $M\left(  C_{1},C_{2}\right)$ and $N\left(  C_{1},C_{2}\right)  $ are
some smooth scalar functions of the first two invariants in Eqs. (\ref{cc}).
For the (linear) vacuum case, $\mathbf{D}=\varepsilon_{0}\mathbf{E}$
and $\mathbf{B}=\mu_{0}\mathbf{H}$, where $\varepsilon_{0}$ and $\mu_{0}$ are
respectively the permittivity and
permeability of empty space; and where, consistent
with Eqs. (\ref{con1}), $\varepsilon_{0}\mu_{0}=c^{-2}$ (so that
$M = 0$, $N = 1/\mu_0 = \varepsilon_0 c^2)$. But in general,
the dependence of $C_1$ and $C_2$ on $\mathbf{B}$ and $\mathbf{E}$ means
that Eqs. (\ref{con1}) are nonlinear. When the constitutive equations take
this form, the other Lorentz invariants in Eqs. (\ref{cc}) are determined
from $C_{1}$ and $C_{2}$ by the formulas,
\begin{align}
C_{3} &  =[\,M\left(  C_{1},C_{2}\right)^2  -\dfrac{1}{c^{2}}%
N\left(  C_{1},C_{2}\right)^2  \,]\,  C_{1}+\dfrac{4}{c^{2}}M\left(
C_{1},C_{2}\right)  N\left(  C_{1},C_{2}\right)  C_{2},\nonumber\\
C_{4} &  =M\left(  C_{1},C_{2}\right)  N\left(  C_{1},C_{2}\right)
C_{1}-[\,  M\left(  C_{1},C_{2}\right)^2  -\dfrac{1}{c^{2}}N\left(
C_{1},C_{2}\right)^2  \,]\,  C_{2},\nonumber\\
C_{5} &  =N\left(  C_{1},C_{2}\right)  C_{1}-2M\left(  C_{1},C_{2}\right)
C_{2},\quad
C_{6}  =M\left(  C_{1},C_{2}\right)  C_{1}+\dfrac{2}{c^{2}}N\left(
C_{1},C_{2}\right)  .\label{c2}%
\end{align}
If the theory is conformally invariant, then the
\textquotedblleft constitutive functions\textquotedblright\ $M$ and $N$
depend only on ${C_{1}}/{C_{2}}$, so that
$M\left(  C_{1},C_{2}\right)  =M_{\mathrm{conf}}\left({C_{1}}/{C_{2}}\right)$
and $N\left(C_{1},C_{2}\right)  =N_{\mathrm{conf}}\left({C_{1}}/{C_{2}}\right)$
\cite{fus/tsi}. Born-Infeld electrodynamics is specified by
\begin{align}
M\left(  C_{1},C_{2}\right)   &  =
\dfrac{C_{2}}{\mu_0 b^2\sqrt{1+\dfrac{c^2}{b^2}C_{1}-\dfrac{c^2}
{b^{4}}C_{2}^{2}}}\,,\label{bi1}\\
N\left(  C_{1},C_{2}\right)   &  =
\dfrac{1}{\mu_0\sqrt{1+\dfrac{c^2}{b^2}C_{1}-\dfrac{c^2}{b^{4}} C_{2}^{2}}}\,,\label{bi2}
\end{align}
where the parameter $b$ is the maximum electric field strength
when the magnetic field is zero.
It follows that Born-Infeld nonlinear electrodynamics has no conformal symmetry.

In covariant notation
\begin{equation}
C_{1}=\dfrac{1}{2}F_{\mu\nu}F^{\mu\nu}\equiv2X,\ \ \ \ \ \ \ \ C_{2}%
=-\dfrac{c}{4}F_{\mu\nu}\tilde{F}^{\mu\nu}\equiv-cY\, \label{xy}%
\end{equation}
where $X$ and $Y$ are introduced here (for later use) for compatibility
with the standard invariant notation. Then we also have
\begin{equation}
G_{\mu\nu}=N\left(  C_{1},C_{2}\right)  F_{\mu\nu}+cM\left(  C_{1}%
,C_{2}\right)  \tilde{F}_{\mu\nu}\,. \label{con2}%
\end{equation}
Taking the Hodge conjugate of Eq. (\ref{con2}), we
represent these equations in the form
\begin{equation}
\binom{G_{\mu\nu}}{\tilde{G}_{\mu\nu}}=\left(
\begin{array}
[c]{cc}%
N\left(  C_{1},C_{2}\right)  & cM\left(  C_{1},C_{2}\right) \\
-cM\left(  C_{1},C_{2}\right)  & N\left(  C_{1},C_{2}\right)
\end{array}
\right)  \binom{F_{\mu\nu}}{\tilde{F}_{\mu\nu}}. \label{ggff}%
\end{equation}
In the space of \textquotedblleft spinors\textquotedblright%
\begin{equation}
\Pi^{F}=\binom{F_{\mu\nu}}{\tilde{F}_{\mu\nu}},\ \ \ \ \ \Pi^{G}=\binom
{G_{\mu\nu}}{\tilde{G}_{\mu\nu}}\,,%
\end{equation}
we then have a kind of quaternionic structure as discussed in
Refs. \cite{cirl04,cir1},%
\begin{equation}
\Pi^{G}=\mathbb{Q}\cdot\Pi^{F},
\end{equation}
where $\mathbb{Q}$ is defined by%
\begin{equation}
\mathbb{Q}=N\left(  C_{1},C_{2}\right)  \sigma_{0}+icM\left(  C_{1}%
,C_{2}\right)  \sigma_{2}\,,%
\end{equation}
and where $\sigma_{0}=I=\left(
\begin{array}
[c]{cc}%
1 & 0\\
0 & 1
\end{array}
\right)  $, $\sigma_{2}=\left(
\begin{array}
[c]{cc}%
0 & -i\\
i & 0
\end{array}
\right) $ are Pauli matrices.

\section{Generalized nonlinear constitutive equations}

Although the constitutive equations (\ref{con1}) are fairly general,
they do not take account of a variety of possibilities, such as
anisotropic media \cite{dmi1}, chiral materials where
derivative terms enter \cite{vin1}, piroelectric and ferromagnetic
materials, and so forth. Therefore we propose to generalize
Eqs. (\ref{con1}) and (\ref{con2}) by introducing the
{\it constitutive tensors\/} $S_{\mu\nu}$, $R_{\mu\nu}^{\rho\sigma}$, and $Q_{\mu\nu}^{\rho\sigma\lambda_{1}\ldots
\lambda_{n}}$, $n = 1,2,3,\,\dots$ We then write
a general nonlinear constitutive equation as follows:
\[
G_{\mu\nu}=S_{\mu\nu}\,+\,R_{\mu\nu}^{\rho\sigma}F_{\rho\sigma}\,+\,Q_{\mu\nu}%
^{\rho\sigma\lambda_{1}}\dfrac{\partial F_{\rho\sigma}}{\partial
x^{\lambda_{1}}}\,+\,Q_{\mu\nu}^{\rho\sigma\lambda_{1}\lambda_{2}}\dfrac{\partial
F_{\rho\sigma}}{\partial x^{\lambda_{1}}\partial x^{\lambda_{2}}}\,+\,\ldots
\]
\begin{equation}
+\, Q_{\mu\nu}^{\rho\sigma\lambda_{1}\ldots
\lambda_{n}}\dfrac{\partial
F_{\rho\sigma}}{\partial x^{\lambda_{1}}\partial x^{\lambda_{2}}\cdots\partial x^{\lambda_n}}\,.
\label{con}%
\end{equation}
This equation, and its superfield analogue
introduced in Sec. 6 below, are the central focus
of the approach we are advocating here. Evidently the
formula (\ref{con}), taken
together with Maxwell's equations, includes all the
possibilities discussed up to now, as well as
new ones, defining the {\it general nonlinear electromagnetic theory}.

Let us discuss the arguments of the constitutive tensors which
are the coefficients in Eq. (\ref{con}). If $S_{\mu\nu}$ were
to depend on $F_{\rho\sigma}$, we could just rewrite it as
a term of the form $R_{\mu\nu}^{\rho\sigma}F_{\rho\sigma}$,
with $R_{\mu\nu}^{\rho\sigma} = S_{\mu\nu}/F_{\rho\sigma}$, and
thus incorporate it into the second term of Eq. (\ref{con}). Hence
we must distinguish $S_{\mu\nu}$ by the absence of any dependence
on $F_{\rho\sigma}$; its arguments (for maximum generality) are the
spacetime coordinates $x^\lambda$. Likewise $R_{\mu\nu}^{\rho\sigma}$
can depend on $x^\lambda$ and on $F_{\kappa\lambda}$, but not on
any derivatives of $F_{\kappa\lambda}$. In general,
$Q_{\mu\nu}^{\rho\sigma\lambda_{1}\ldots
\lambda_{n}}$ depends on $x^\lambda$
together with derivatives of $F_{\kappa\lambda}$ of $n$th order
or less.

Now one can require $R_{\mu\nu}^{\rho\sigma}$ and
$Q_{\mu\nu}^{\rho\sigma\lambda_{1}\ldots \lambda_{n}}$
to depend directly on $x$, $F_{\kappa\lambda}$,
and derivatives of appropriate orders, as {\it local\/}
functions that take into account only their values at $x$.
But one may also consider constitutive tensors that
depend on the field strengths and their derivatives
via functionals that take account of their values at
different spacetime points, or that involve integrals of
functions of field strengths and their derivatives over regions
of spacetime. Thus we also have the capability of describing
a variety of nonlocal effects with this approach.

As we impose Lorentz covariance on the constitutive equations,
the constitutive tensors will depend on the fields through
the invariants $X$ and $Y$, as follows: $S_{\mu\nu}$ is a constant
independent of $X$ and $Y$, while
\begin{equation}
R_{\mu\nu}^{\rho\sigma}=R_{\mu\nu}^{\rho\sigma}\left(  X,Y\right)
,\ \ \ \ \ Q_{\mu\nu}^{\rho\sigma\lambda_{1}\ldots\lambda_{n}}=Q_{\mu\nu
}^{\rho\sigma\lambda_{1}\ldots\lambda_{n}}\left(  X,Y,\ldots\right)  ,
\end{equation}
where \textquotedblleft\ldots\textquotedblright\ denotes invariant derivatives of the
invariants $X$, $Y$ up to $n$th order. Obviously $S_{\mu\nu}$ is
antisymmetric, $R_{\mu\nu}^{\rho\sigma}$ is antisymmetric in its upper and lower
indices separately, and the $Q_{\mu\nu}^{\rho\sigma\lambda_{1}\ldots\lambda_{n}}$
are antisymmetric in their upper and first two lower indices; with respect to the
$\lambda_{i}$, they are symmetric Lorentz tensors.

Let us write some familiar examples in this form. For the simplest
vacuum case, we have
\begin{equation}
S_{\mu\nu}=0,\ \ \ R_{\mu\nu}^{\rho\sigma}=\mu_{0}^{-1}\delta_{\lbrack\mu
}^{\rho}\delta_{\nu]}^{\sigma},\ \ \ Q_{\mu\nu}^{\rho\sigma\lambda_{1}%
\ldots\lambda_{n}}=0,
\end{equation}
where the square
brackets denote antisymmetrization with a factor of $1/2$; i.e., $x_{[\mu\nu
]}\equiv\left(  x_{\mu\nu}-x_{\nu\mu}\right) /2$.
The only nonvanishing constitutive tensor $R_{\mu\nu}^{\rho\sigma}$ is
``diagonal.'' For Born-Infeld nonlinear electrodynamics, we have%
\begin{align}
& S_{\mu\nu}  =0,\nonumber\\
& R_{\mu\nu}^{\rho\sigma}  =\dfrac{\delta_{\lbrack\mu}^{\rho}\delta_{\nu
]}^{\sigma}-{\dfrac{c^2}{b^2}}Y\varepsilon_{\lbrack\mu\nu]\lambda\delta}\eta^{\lambda\rho}%
\eta^{\delta\sigma}}{\mu_0\sqrt{1+2\dfrac{c^2}{b^2}X-\dfrac{c^4}{b^4}Y^{2}}},\\
& Q_{\mu\nu}^{\rho\sigma\lambda_{1}\ldots\lambda_{n}}  =0.\nonumber
\end{align}
For an anisotropic medium with tensorial permeability $\varepsilon_{ij}$ and
permittivity $\mu_{ij}$, the
constitutive equations are $\mathbf{D}_{i}=\varepsilon_{ij}\mathbf{E}_{j}%
$, $\mathbf{B}=\mu_{ij}\mathbf{H}_{j}$. This case is not described by Eq.
(\ref{con2}), but the corresponding
constitutive tensor $R_{\mu\nu}^{\rho\sigma}$ is
easily calculated; while again, $S_{\mu\nu}$ and the $Q_{\mu\nu}%
^{\rho\sigma\lambda_{1}\ldots\lambda_{n}}$ vanish. The case $S_{\mu\nu}\neq 0$
describes piroelectric and ferromagnetic materials.

When the equations of motion for the nonlinear theory derive from a Lagrangian
$L\left(  X,Y\right)$ which is a scalar
function of the invariants $X$ and $Y$ but does not depend
on their derivatives, then from the usual definitions
together with Eq. (\ref{con2}) we have
\begin{equation}
G_{\mu\nu}=-\dfrac{\partial L\left(  X,Y\right)  }{\partial X}F_{\mu\nu}%
-\dfrac{\partial L\left(  X,Y\right)  }{\partial Y}\tilde{F}_{\mu\nu}.
\label{gf}%
\end{equation}
Comparing (\ref{con}) and (\ref{gf}) gives us the
constitutive tensors,
\begin{align}
& S_{\mu\nu}   =0,\nonumber\\
& R_{\mu\nu}^{\rho\sigma}   =-\dfrac{\partial L\left(  X,Y\right)  }{\partial
X}\delta_{\lbrack\mu}^{\rho}\delta_{\nu]}^{\sigma}-\dfrac{\partial L\left(
X,Y\right)  }{\partial Y}\varepsilon_{\lbrack\mu\nu]\lambda\delta}%
\eta^{\lambda\rho}\eta^{\delta\sigma},\label{rl}\\
& Q_{\mu\nu}^{\rho\sigma\lambda_{1}\ldots\lambda_{n}}   =0.\nonumber
\end{align}
In this case the functions $M$ and $N$ in Eqs. (\ref{con1}) are%
\begin{equation}
N_{L}\left(  X,Y\right)  =-\dfrac{\partial L\left(  X,Y\right)  }{\partial
X},\ \ \ \ \ \ \ \ M_{L}\left(  X,Y\right)  =-\dfrac{1}{c}\dfrac{\partial
L\left(  X,Y\right)  }{\partial Y}. \label{nm}%
\end{equation}
So for the
``constitutive functions'' $M$ and $N$ to describe a Lagrangian
theory of nonlinear electrodynamics, we need
\begin{equation}
\dfrac{\partial N_{L}\left(  X,Y\right)  }{\partial Y}=c\dfrac{\partial
M_{L}\left(  X,Y\right)  }{\partial X}.
\end{equation}
We see that dissipative, non-Lagrangian theories are naturally
included in the current framework.

\section{Duality transformations}

Next consider the duality transformation $\delta$
transforming $F_{\mu\nu}$ to
$\tilde{G}_{\mu\nu}$ and $G_{\mu\nu}$ to $\tilde{F}_{\mu\nu}$ \cite{gib/has}:
\begin{equation}
\delta F_{\mu\nu}=\tilde{G}_{\mu\nu},\ \ \ \ \ \delta G_{\mu\nu}=\tilde
{F}_{\mu\nu}\,. \label{dua}%
\end{equation}
The self-duality $(+)$ or antiself-duality $(-)$ condition is defined by%
\begin{equation}
F_{\mu\nu}   = \epsilon\,\tilde{G}_{\mu\nu}\,,\quad \epsilon = \pm 1, \label{sd1}
%
%
\end{equation}
which establishes the main relation of a self-dual theory,%
\begin{equation}
F_{\mu\nu}\tilde{F}^{\mu\nu}=G_{\mu\nu}\tilde{G}^{\mu\nu}%
\end{equation}
(which is equivalent to $\mathbf{D\cdot H}=\mathbf{B\cdot E})$.
Making reference to Eq. \!(\ref{con}), we have the corresponding (anti)self-duality
conditions for the constitutive tensors $R_{\mu\nu}^{\rho\sigma}$
(with $S_{\mu\nu}=0$, $Q_{\mu\nu}^{\rho
\sigma\lambda_{1}\ldots\lambda_{n}}=0$),
\begin{equation}
R_{\mu\nu}^{\rho\sigma}\varepsilon_{\rho\sigma\lambda\delta}\eta
^{\lambda\lbrack\mu}\eta^{\delta\nu]}\,=\,2\epsilon\,.
\end{equation}
The further requirement that a solution of (\ref{sd1}) also
obey the (anti)self-duality condition
$F_{\mu\nu}   = \epsilon\,\tilde{F}^{\mu\nu}$ implies that $X = Y$,
but we see these two conditions as fundamentally distinct.
We can obtain equations of motion in this framework by the method of
Ref. \cite{gib/has}.

More generally, the finite duality transformation are given by
\begin{align}
F_{\mu\nu}^{\,\prime}  &  \,=\,aF_{\mu\nu}+ b\tilde{G}_{\mu\nu},\nonumber\\
G_{\mu\nu}^{\,\prime}  &  \,=\,eG_{\mu\nu}+ f\tilde{F}_{\mu\nu},
\end{align}
where the determinant $af-be=1$. If we
use the constitutive equations (\ref{con2}) and their Hodge
conjugates, we can write
\begin{align}
F_{\mu\nu}^{\,\prime}  &  =U_{\mu\nu}^{\rho\sigma}F_{\rho\sigma},\nonumber\\
G_{\mu\nu}^{\,\prime}  &  =V_{\mu\nu}^{\rho\sigma}G_{\rho\sigma},
\end{align}
where the ``dual tensors'' $U_{\mu\nu}^{\rho\sigma}$ and $V_{\mu\nu}^{\rho\sigma}$
take the form%
\begin{align}
U_{\mu\nu}^{\rho\sigma}  &  = [ \, a-bcM\left(  C_{1},C_{2}\right) \,]
\delta_{\mu}^{\rho}\delta_{\nu}^{\sigma}+\dfrac{1}{2}\,bN\left(  C_{1}%
,C_{2}\right)  \varepsilon_{\mu\nu\lambda\delta}\eta^{\lambda\rho}\eta
^{\delta\sigma},\nonumber\\
V_{\mu\nu}^{\rho\sigma}  &  =\left[ \,e+\dfrac{fcM\left(  C_{1},C_{2}\right)
}{N\left(  C_{1},C_{2}^2\right)  +c^{2}M\left(  C_{1},C_{2}^2\right)
}\,\right]  \delta_{\mu}^{\rho}\delta_{\nu}^{\sigma}\nonumber\\
&  +\dfrac{1}{2}\dfrac{N\left(  C_{1},C_{2}\right)  }{N\left(  C_{1}%
,C_{2}\right)^2  +c^{2}M \left(  C_{1},C_{2}\right)^2  }\,\varepsilon_{\mu
\nu\lambda\delta}\eta^{\lambda\rho}\eta^{\delta\sigma}.
\end{align}

\section{Supersymmetric electrodynamics}

Now we turn to a geometric superfield formulation of classical electrodynamics
and the constitutive equations in superspace. We start by fixing some further
notation, following mostly Ref. \cite{wes/bag}. The $N=1$ four-dimensional
superspace is described by
coordinates $z^{M}=\left\{  x^{\mu},\theta^{\alpha},\bar{\theta}%
^{\dot{\alpha}}\right\}  $, where we introduce the unifying index $M=\left\{
\mu,\alpha,\dot{\alpha}\right\}  $ and $\theta^{\alpha},\bar{\theta}%
^{\dot{\alpha}}$ ($\alpha,\dot{\alpha}=1,2$) are additional complex Grassmann
coordinates (two-components Majorana spinors). The transformations of ($N=1$,
$D=4$) supersymmetry are given by%
\begin{align}
\tilde{x}^{\mu}  &  =x^{\mu}+i\lambda^{\alpha}\sigma_{\alpha\dot{\alpha}}%
^{\mu}\bar{\theta}^{\dot{\alpha}}-i\theta^{\alpha}\sigma_{\alpha\dot{\alpha}%
}^{\mu}\lambda^{\dot{\alpha}},\nonumber\\
\tilde{\theta}^{\alpha}  &  =\theta^{\alpha}+\lambda^{\alpha}%
,\ \ \ \ \ \widetilde{\bar{\theta}}^{\,\dot{\alpha}}=\bar{\theta}^{\dot{\alpha}%
}+\lambda^{\dot{\alpha}}\,, \label{s2}%
\end{align}
where $\lambda^{\alpha}$ and $\lambda^{\dot{\alpha}}$ are constant Grassmann
spinors. The transformations in Eqs. (\ref{s2}) are generated by
supercharges%
\begin{align}
\mathrm{Q}_{\alpha}  &  =-i\dfrac{\partial}{\partial\theta^{\alpha}}%
+\sigma_{\alpha\dot{\alpha}}^{\mu}\bar{\theta}^{\dot{\alpha}}\dfrac{\partial
}{\partial x^{\mu}}\,,\nonumber\\
\mathrm{\bar{Q}}_{\dot{\alpha}}  &  =i\dfrac{\partial}{\partial\bar{\theta
}^{\dot{\alpha}}}+\theta^{\alpha}\sigma_{\alpha\dot{\alpha}}^{\mu}%
\dfrac{\partial}{\partial x^{\mu}}\,,\label{supercharges}
\end{align}
with
\begin{equation}
\left\{  \mathrm{Q}_{\alpha},\mathrm{\bar{Q}}_{\dot{\alpha}}\right\}
=2i\sigma_{\alpha\dot{\alpha}}^{\mu}\dfrac{\partial}{\partial x^{\mu}}\,,
\label{qq}%
\end{equation}
where the $\sigma_{\alpha\dot{\alpha}}^{\mu}=\left(
I,\vec{\sigma}\right)  _{\alpha\dot{\alpha}}$ are Pauli matrices. Then
$\tilde{z}^{M}=\exp\,\left[\,i\left(  \lambda^{\alpha}\mathrm{Q}%
_{\alpha}+\mathrm{\bar{Q}}_{\dot{\alpha}}\lambda^{\dot{\alpha}}\,\right)
\right]\!z^{M}$. Defining
\begin{align}
\mathrm{D}_{\mu}  &  =\dfrac{\partial}{\partial x^{\mu}}\,,\nonumber\\
\mathrm{D}_{\alpha}  &  =\dfrac{\partial}{\partial\theta^{\alpha}}%
-i\sigma_{\alpha\dot{\alpha}}^{\mu}\bar{\theta}^{\dot{\alpha}}\dfrac{\partial
}{\partial x^{\mu}}\,,\nonumber\\
\mathrm{\bar{D}}_{\dot{\alpha}}  &  =-\dfrac{\partial}{\partial\bar{\theta
}^{\dot{\alpha}}}+i\theta^{\alpha}\sigma_{\alpha\dot{\alpha}}^{\mu}%
\dfrac{\partial}{\partial x^{\mu}}\,,\label{derivatives}
\end{align}
we have
\begin{equation}
\left\{  \mathrm{D}_{\alpha},\mathrm{\bar{D}}_{\dot{\alpha}}\right\}
=2i\sigma_{\alpha\dot{\alpha}}^{\mu}\dfrac{\partial}{\partial x^{\mu}}\,.
\label{dd}%
\end{equation}
The (anti)commutators other than those of Eqs. (\ref{qq}) and (\ref{dd}) vanish.

Now a general superfield $\Phi\left(  x,\theta,\bar{\theta}\right)$ written
as a
function of nilpotent Grassmann variables $\theta^{\alpha}$ and $\bar{\theta
}^{\dot{\alpha}}$ can be expanded as a finite series with respect to them.
Its components are ordinary and spinorial functions, members of the corresponding
supermultiplet, that are mixed by the (infinitesimal) supersymmetry transformations%
\begin{equation}
\delta\Phi\left(  x,\theta,\bar{\theta}\right)  =i\left(  \lambda^{\alpha
}\mathrm{Q}_{\alpha}+\mathrm{\bar{Q}}_{\dot{\alpha}}\lambda^{\dot{\alpha}%
}\right)  \Phi\left(  x,\theta,\bar{\theta}\right)  .
\end{equation}
For further details, see Ref. \cite{wes/bag}.

The abelian gauge field $A_{\mu}\left(  x\right)  $ is a component of a gauge
superfield vector multiplet $\mathsf{V}\left(  x,\theta,\bar{\theta}\right)
=\mathsf{V}^{+}(  x,\theta,\bar{\theta})$, where $+$ denotes super-Hermitian conjugation. The supergauge transformations are given by
\begin{equation}
\mathsf{\tilde{V}}\left(  x,\theta,\bar{\theta}\right)  =\mathsf{V}\left(
x,\theta,\bar{\theta}\right)  +\dfrac{i}{2}\left(  \Lambda\left(
x,\theta,\bar{\theta}\right)  -\Lambda^{+}\left(  x,\theta,\bar{\theta
}\right)  \right)  , \label{sg}%
\end{equation}
where $\Lambda\left(  x,\theta,\bar{\theta}\right)  $ is chiral superfield
parameter satisfying $\mathrm{D}_{\alpha}\Lambda\left(  x,\theta,\bar{\theta
}\right)  =0$ and $\mathrm{\bar{D}}_{\dot{\alpha}}\Lambda^{+}\left(
x,\theta,\bar{\theta}\right)  =0$. That is, defining $x_{L,R}^{\mu}=x^{\mu
}\pm i\theta^{\alpha}\sigma_{\alpha\dot{\alpha}}^{\mu}\bar{\theta}%
^{\dot{\alpha}}$, $\Lambda$ and $\Lambda^{+}$ is each actually a function of
two variables, $\Lambda\left(  x,\theta,\bar{\theta}\right)  =\Upsilon\left(
x_{L},\theta\right)  $, and $\Lambda^{+}\left(  x,\theta,\bar{\theta}\right)
=\Upsilon^{+}\left(  x_{R},\bar{\theta}\right)  $. In the Wess-Zumino gauge.
half of the component fields are
gauged away using supergauge transformations (\ref{sg});
so that $\mathsf{V}%
\left(  x,\theta,\bar{\theta}\right)  $ takes the form,
\begin{align}
\mathsf{V}_{WZ}\left(  x,\theta,\bar{\theta}\right)   &  =-\theta^{\alpha
}\sigma_{\alpha\dot{\alpha}}^{\mu}\bar{\theta}^{\dot{\alpha}}A_{\mu}\left(
x\right)  -i\bar{\theta}_{\dot{\alpha}}\bar{\theta}^{\dot{\alpha}}%
\theta^{\alpha}\psi_{\alpha}\left(  x\right) \nonumber\\
&  +i\theta^{\alpha}\theta_{\alpha}\bar{\theta}_{\dot{\alpha}}\bar{\psi}%
^{\dot{\alpha}}\left(  x\right)  +\dfrac{1}{2}\theta^{\alpha}\theta_{\alpha
}\bar{\theta}_{\dot{\alpha}}\bar{\theta}^{\dot{\alpha}}D\left(  x\right)  ,
\label{vwz}%
\end{align}
where $\psi_{\alpha}\left(  x\right)  $ is a Majorana fermion field (photino),
and $D\left(  x\right)  $ is an auxiliary field
which vanishes on-shell.
%
%

If $\Psi\left(  x,\theta,\bar{\theta}\right)$ is a matter
superfield, then the supergauge transformations of $\Psi$
are given by,
\begin{equation}
\tilde{\Psi}\left(  x,\theta,\bar{\theta}\right)  =\exp\left[  -\dfrac
{ie}{\hbar c}\left(  \Lambda\left(  x,\theta,\bar{\theta}\right)  +\Lambda
^{+}\left(  x,\theta,\bar{\theta}\right)  \right)  \right]  \Psi\left(
x,\theta,\bar{\theta}\right)  . \label{sgm}%
\end{equation}
Together, Eqs. (\ref{sg}) and (\ref{sgm}) provide the full set of supergauge
transformations of the abelian $N=1$ gauge theory.

Next let us introduce the gauge superpotential superfield
$\mathsf{A}_{M}\left(  x,\theta,\bar{\theta}\right)$, a superconnection
on the supermanifold, corresponding to
the superderivatives $\mathrm{D}_{M}$. Then the covariant superderivatives
(covariant with respect to the supergauge
transformations (\ref{sg})-(\ref{sgm})) are
\begin{equation}
\nabla_{M}=\mathrm{D}_{M}+\dfrac{ie}{\hbar c}\mathsf{A}_{M}\left(
x,\theta,\bar{\theta}\right)  .
\end{equation}

The requirement that the covariant superderivatives acting on the matter
superfield transform as superfields themselves via Eq. (\ref{sgm}) leads to

\begin{align}
\mathsf{\tilde{A}}_{\mu}\left(  x,\theta,\bar{\theta}\right)   &
=\mathsf{A}_{\mu}\left(  x,\theta,\bar{\theta}\right)  +\dfrac{1}{2}%
\mathrm{D}_{\mu}\left(  \Lambda\left(  x,\theta,\bar{\theta}\right)
+\Lambda^{+}\left(  x,\theta,\bar{\theta}\right)  \right)  ,\nonumber\\
\mathsf{\tilde{A}}_{\alpha}\left(  x,\theta,\bar{\theta}\right)   &
=\mathsf{A}_{\alpha}\left(  x,\theta,\bar{\theta}\right)  +\dfrac{1}%
{2}\mathrm{D}_{\alpha}\Lambda^{+}\left(  x,\theta,\bar{\theta}\right)  ,\nonumber\\
\widetilde{\mathsf{\bar{A}}}_{\dot{\alpha}}\left(  x,\theta,\bar{\theta
}\right)   &  =\mathsf{\bar{A}}_{\dot{\alpha}}\left(  x,\theta,\bar{\theta
}\right)  +\dfrac{1}{2}\mathrm{\bar{D}}_{\dot{\alpha}}\Lambda\left(
x,\theta,\bar{\theta}\right)  .
\end{align}
These relations can be satisfied identically by choosing
\begin{equation}
\mathsf{A}_{\mu}\left(  x,\theta,\bar{\theta}\right)  =\left(
i\mathrm{D}_{\mu}-\dfrac{1}{2}\mathrm{D}_{\alpha}\sigma_{\mu}^{\alpha
\dot{\alpha}}\mathrm{\bar{D}}_{\dot{\alpha}}\right)  \mathsf{V}\left(
x,\theta,\bar{\theta}\right)\label{a1}
\end{equation}
and
\begin{align}
\mathsf{A}_{\alpha}\left(  x,\theta,\bar{\theta}\right)   &  =i\mathrm{D}%
_{\alpha}\mathsf{V}\left(  x,\theta,\bar{\theta}\right)\,,\nonumber\\
\mathsf{\bar{A}}_{\dot{\alpha}}\left(  x,\theta,\bar{\theta}\right)   &
=i\mathrm{\bar{D}}_{\dot{\alpha}}\mathsf{V}\left(  x,\theta,\bar{\theta
}\right)\,, \label{a2}%
\end{align}
where $\mathsf{V}\left(  x,\theta,\bar{\theta}\right)  $ is a prepotential of
the $N=1$ abelian gauge theory. From Eqs. (\ref{a1})--(\ref{a2})
the relation between the vector and spinor covariant superderivatives follows,%
\begin{equation}
\nabla_{\mu}=\dfrac{i}{4}\,\sigma_{\mu}^{\alpha\dot{\alpha}}\left\{
\nabla_{\alpha},\bar{\nabla}_{\dot{\alpha}}\right\}  , \label{ddd}%
\end{equation}
using the relation between covariant superderivatives and superderivatives,%
\begin{align}
\nabla_{\alpha}  &  =e^{\ \tfrac{e}{\hbar c}\mathsf{V}\left(  x,\theta
,\bar{\theta}\right)  }\mathrm{D}_{\alpha}e^{-\tfrac{e}{\hbar c}%
\mathsf{V}\left(  x,\theta,\bar{\theta}\right)  },\nonumber\\
\bar{\nabla}_{\dot{\alpha}}  &  =e^{-\tfrac{e}{\hbar c}\mathsf{V}\left(
x,\theta,\bar{\theta}\right)  }\mathrm{\bar{D}}_{\dot{\alpha}}e^{\ \tfrac
{e}{\hbar c}\mathsf{V}\left(  x,\theta,\bar{\theta}\right)  }\,.%
\end{align}

Now one constructs the corresponding (anti)commutators of the
covariant superderivatives, introducing the gauge invariant torsion $\mathsf{T}%
_{MN}^{\ \ \ \ K}$ and the superfield strength $\mathsf{F}_{MN}\left(
x,\theta,\bar{\theta}\right)  $, as follows. Let $\{\ \ ]$ denote
the ``mixed commutator''---the
anticommutator when both entries are odd,
the commutator for other combinations. Then we have,
\begin{equation}
\{\nabla_{M},\nabla_{N}]=i\mathsf{T}_{MN}^{\ \ \ \ K}\nabla_{K}+i\mathsf{F}%
_{MN}\left(  x,\theta,\bar{\theta}\right).
\end{equation}
From Eq. (\ref{ddd}), it follows
that the only nonvanishing components of the torsion are
\begin{equation}
\mathsf{T}_{\alpha\dot{\alpha}}^{\ \ \ \ \mu}=2\sigma_{\alpha\dot{\alpha}%
}^{\mu}.
\end{equation}
Thus in the $N=1$ abelian theory, the torsion constraints are the same as in
flat $N=1$ superspace.

For the superfield strength, it follows from
Eq. (\ref{ddd}) that all the components for which both indices are fermionic vanish;
i.e.,
\begin{equation}
\mathsf{F}_{\alpha\beta}\left(  x,\theta,\bar{\theta}\right)  =\mathsf{F}%
_{\alpha\dot{\beta}}\left(  x,\theta,\bar{\theta}\right)  =\mathsf{F}%
_{\dot{\alpha}\dot{\beta}}\left(  x,\theta,\bar{\theta}\right)  =0.
\end{equation}
These constraints are called ``representation preserving,'' because they allow
one to introduce chiral and anti-chiral superfields which survive in the presence
of nonzero gauge coupling. Then the lowest dimensional surviving superfield
strengths are mixed spin-vector (odd valued) superfields,
\begin{align}
\mathsf{F}_{\alpha\mu}\left(  x,\theta,\bar{\theta}\right)   &  =-i\left[
\nabla_{\alpha},\nabla_{\mu}\right]  =\mathrm{D}_{\alpha}\mathsf{A}_{\mu
}\left(  x,\theta,\bar{\theta}\right)  -\mathrm{D}_{\mu}\mathsf{A}_{\alpha
}\left(  x,\theta,\bar{\theta}\right)  ,\nonumber\\
\mathsf{\bar{F}}_{\dot{\alpha}\mu}\left(  x,\theta,\bar{\theta}\right)   &
=-i\left[  \bar{\nabla}_{\dot{\alpha}},\nabla_{\mu}\right]  =\mathrm{\bar{D}%
}_{\dot{\alpha}}\mathsf{A}_{\mu}\left(  x,\theta,\bar{\theta}\right)
-\mathrm{D}_{\mu}\mathsf{\bar{A}}_{\dot{\alpha}}\left(  x,\theta,\bar{\theta
}\right).
\end{align}
These are the actual super analogues of the field strength $F_{\mu\nu}$
in ordinary electromagnetism.

From Eqs. (\ref{a1})--(\ref{a2}), the manifest form of
the superfield strengths in terms of the prepotential is
\begin{align}
\mathsf{F}_{\alpha\mu}\left(  x,\theta,\bar{\theta}\right)   &  =-\dfrac{1}%
{2}\mathrm{D}_{\alpha}\mathrm{D}_{\beta}\sigma_{\mu}^{\beta\dot{\beta}%
}\mathrm{\bar{D}}_{\dot{\beta}}\mathsf{V}\left(  x,\theta,\bar{\theta}\right)
,\nonumber\\
\mathsf{\bar{F}}_{\dot{\alpha}\mu}\left(  x,\theta,\bar{\theta}\right)   &
=-\dfrac{1}{2}\mathrm{\bar{D}}_{\dot{\alpha}}\mathrm{\bar{D}}_{\dot{\beta}%
}\sigma_{\mu}^{\dot{\beta}\beta}\mathrm{D}_{\beta}\mathsf{V}\left(
x,\theta,\bar{\theta}\right)  . \label{fd2}%
\end{align}
The superfield strengths can be expressed in terms of a chiral spinor
superfield that depends on only one spinorial coordinate, as follows:%
\begin{align}
\mathsf{F}_{\alpha\mu}\left(  x,\bar{\theta}\right)   &  =-i\varepsilon
_{\alpha\beta}\sigma_{\mu}^{\beta\dot{\beta}}\mathsf{\bar{W}}_{\dot{\beta}%
}\left(  x,\bar{\theta}\right)  ,\nonumber\\
\mathsf{\bar{F}}_{\dot{\alpha}\mu}\left(  x,\theta\right)   &  =-i\varepsilon
_{\dot{\alpha}\dot{\beta}}\bar{\sigma}_{\mu}^{\dot{\beta}\beta}\mathsf{W}%
_{\beta}\left(  x,\theta\right)  , \label{f2}%
\end{align}
where%
\begin{align}
\mathsf{W}_{\beta}\left(  x,\theta\right)   &  =\dfrac{1}{2}\mathrm{\bar{D}%
}_{\dot{\alpha}}\mathrm{\bar{D}}^{\dot{\beta}}\mathrm{D}_{\beta}%
\mathsf{V}\left(  x,\theta,\bar{\theta}\right)  ,\ \ \ \ \ \mathrm{\bar{D}%
}_{\dot{\alpha}}\mathsf{W}_{\beta}\left(  x,\theta\right)  =0,\nonumber\\
\mathsf{\bar{W}}_{\dot{\beta}}\left(  x,\bar{\theta}\right)   &  =\dfrac{1}%
{2}\mathrm{D}^{\alpha}\mathrm{D}_{\beta}\mathrm{\bar{D}}_{\dot{\beta}%
}\mathsf{V}\left(  x,\theta,\bar{\theta}\right)  ,\ \ \ \ \ \mathrm{D}%
_{\alpha}\mathsf{\bar{W}}_{\dot{\beta}}\left(  x,\bar{\theta}\right)  =0.
\label{wd2}%
\end{align}
These chiral superfields satisfy an additional constraint, $\,\mathrm{D}^{\alpha
}\mathsf{W}_{\alpha}\left(  x,\theta\right)  =\mathrm{\bar{D}}_{\dot{\alpha}%
}\mathsf{\bar{W}}^{\dot{\alpha}}\left(  x,\bar{\theta}\right)  $.

Using the component expansion of $\mathsf{V}\left(  x,\theta,\bar{\theta
}\right)$ in the Wess-Zumino gauge (\ref{vwz}), we obtain
\begin{align}
\mathsf{W}_{\alpha}\left(  x,\theta\right)   &  =-i\psi_{\alpha}\left(
x\right)  +\left(  \varepsilon_{\alpha\gamma}D\left(  x\right)  -\dfrac{i}%
{2}\sigma_{\alpha\dot{\alpha}}^{\mu}\varepsilon^{\dot{\alpha}\dot{\beta}}%
\bar{\sigma}_{\dot{\beta}\gamma}^{\nu}F_{\mu\nu}\left(  x\right)  \right)
\theta^{\gamma}-\theta^{\beta}\theta_{\beta}\sigma_{\alpha\dot{\alpha}}^{\mu
}\partial_{\mu}\bar{\psi}^{\dot{\alpha}}\left(  x\right)  ,\nonumber\\
\mathsf{\bar{W}}_{\dot{\alpha}}\left(  x,\bar{\theta}\right)   &  =i\bar{\psi
}_{\dot{\alpha}}\left(  x\right)  +\left(  \varepsilon_{\dot{\alpha}%
\dot{\gamma}}D\left(  x\right)  +\dfrac{i}{2}\bar{\sigma}_{\dot{\alpha}\alpha
}^{\mu}\varepsilon^{\alpha\beta}\sigma_{\beta\dot{\gamma}}^{\nu}F_{\mu\nu
}\left(  x\right)  \right)  \bar{\theta}^{\dot{\gamma}}+\bar{\theta}%
_{\dot{\beta}}\bar{\theta}^{\dot{\beta}}\bar{\sigma}_{\dot{\alpha}\alpha}%
^{\mu}\partial_{\mu}\psi^{\alpha}\left(  x\right). \label{w2}%
\end{align}
From Eqs. \!(\ref{f2}), it follows that the role the gauge invariants $X$
and $Y$ in Eqs. \!(\ref{xy}) played for nonlinear electromagnetism is
now played by the superfield invariants,
\begin{align}
\mathsf{X}\left(  x,\theta\right)   &  =\dfrac{1}{4}\mathsf{\bar{F}}%
_{\dot{\alpha}\mu}\left(  x,\theta\right)  \mathsf{\bar{F}}^{\dot{\alpha}\mu
}\left(  x,\theta\right)  =\mathsf{W}^{\alpha}\left(  x,\theta\right)
\mathsf{W}_{\alpha}\left(  x,\theta\right)  ,\nonumber\\
\mathsf{Y}\left(  x,\bar{\theta}\right)   &  =\dfrac{1}{4}\mathsf{F}%
^{\alpha\mu}\left(  x,\bar{\theta}\right)  \mathsf{F}_{\alpha\mu}\left(
x,\bar{\theta}\right)  =\mathsf{\bar{W}}_{\dot{\alpha}}\left(  x,\bar{\theta
}\right)  \mathsf{\bar{W}}^{\dot{\alpha}}\left(  x,\bar{\theta}\right)  .
\label{yy}%
\end{align}
Applying the component expansions (\ref{w2}), we observe that
\begin{align}
\mathsf{X}\left(  x,\theta\right)   &  =\,\ldots\,+\,\theta^{\alpha}\theta_{\alpha
}\left(  X-iY\right)  ,\nonumber\\
\mathsf{Y}\left(  x,\bar{\theta}\right)   &  =\,\ldots\,+\,\bar{\theta}_{\dot
{\alpha}}\bar{\theta}^{\dot{\alpha}}\left(  X+iY\right)  ,
\end{align}
which, after integration over the Grassmann coordinates, will give the correct
contributions of $X$ and $Y$ to the Lagrangian.

\section{Supersymmetric constitutive equations}

Finally we are ready to consider the $N=1$ supersymmetric theory in
``supermedia,'' with the goal of
obtaining the ``super'' version of the constitutive
equations (\ref{con}).

By analogy with the case of nonlinear
electromagnetism, we introduce the superfield strengths in
media $\mathsf{G}^{\alpha\mu}\left(  x,\bar{\theta}\right)  $ and
$\mathsf{\bar{G}}_{\dot{\alpha}\mu}\left(  x,\theta\right)  $, without
expressing $\mathsf{G}$ or $\mathsf{\bar{G}}$ via any prepotential. Rather we
assume them to have a
representation similar to that of Eqs. \!(\ref{f2}),%
\begin{align}
\mathsf{G}_{\alpha\mu}\left(  x,\bar{\theta}\right)   &  =-i\varepsilon
_{\alpha\beta}\sigma_{\mu}^{\beta\dot{\beta}}\mathsf{\bar{W}}_{\dot{\beta}%
}^{G}\left(  x,\bar{\theta}\right)  ,\nonumber\\
\mathsf{\bar{G}}_{\dot{\alpha}\mu}\left(  x,\theta\right)   &  =-i\varepsilon
_{\dot{\alpha}\dot{\beta}}\bar{\sigma}_{\mu}^{\dot{\beta}\beta}\mathsf{W}%
_{\beta}^{G}\left(  x,\theta\right). \label{g2}%
\end{align}
Here the chiral superfields in media $\mathsf{W}_{\beta}^{G}\left(
x,\theta\right)  $ and $\mathsf{\bar{W}}_{\dot{\beta}}^{G}\left(
x,\bar{\theta}\right)  $ likewise are not expressed through a
prepotential superfield, but nevertheless have a component
 expansion similar to that of Eqs. \!(\ref{w2}),
\begin{align}
\mathsf{W}_{\alpha}^{G}\left(  x,\theta\right)   &  =-i\psi_{\alpha}%
^{G}\left(  x\right)  +\left(  \varepsilon_{\alpha\gamma}D^{G}\left(
x\right)  -\dfrac{i}{2}\sigma_{\alpha\dot{\alpha}}^{\mu}\varepsilon
^{\dot{\alpha}\dot{\beta}}\bar{\sigma}_{\dot{\beta}\gamma}^{\nu}G_{\mu\nu
}\left(  x\right)  \right)  \theta^{\gamma}-\theta^{\beta}\theta_{\beta}%
\sigma_{\alpha\dot{\alpha}}^{\mu}\partial_{\mu}\bar{\psi}^{G\dot{\alpha}%
}\left(  x\right)  ,\nonumber\\
\mathsf{\bar{W}}_{\dot{\alpha}}^{G}\left(  x,\bar{\theta}\right)   &
=i\bar{\psi}_{\dot{\alpha}}^{G}\left(  x\right)  +\left(  \varepsilon
_{\dot{\alpha}\dot{\gamma}}D^{G}\left(  x\right)  +\dfrac{i}{2}\bar{\sigma
}_{\dot{\alpha}\alpha}^{\mu}\varepsilon^{\alpha\beta}\sigma_{\beta\dot{\gamma
}}^{\nu}G_{\mu\nu}\left(  x\right)  \right)  \bar{\theta}^{\dot{\gamma}}%
+\bar{\theta}_{\dot{\beta}}\bar{\theta}^{\dot{\beta}}\bar{\sigma}_{\dot
{\alpha}\alpha}^{\mu}\partial_{\mu}\psi^{G\alpha}\left(  x\right)  ,
\label{wg2}%
\end{align}
where $G_{\mu\nu}\left(  x\right)  $ satisfies Eq. \!(\ref{max4}).

In writing the analogue of the constitutive
equations (\ref{con}), it is obviously
much more convenient to deal with variables having only
one spinor index (i.e., the chiral superfields) than
with those having spin-vector
indices (i.e., the superfield strengths). Furthermore, these are related
to each other by constant matrices, through Eqs. \!(\ref{f2}) and (\ref{g2}).
Therefore we formulate the superconstitutive equations in terms of
the chiral superfields, as follows:
\begin{align}
\mathsf{W}_{\alpha}^{G}   &  =\mathsf{S}_{\alpha
}  +\mathsf{R}_{\alpha}^{\beta}
\mathsf{W}_{\beta}+\mathsf{Q}_{\alpha}^{\beta_1\gamma
}  \mathrm{D}_{\beta_1}\mathsf{W}_{\gamma}+
\mathsf{Q}_{\alpha}^{\beta_1\beta_2\gamma}
\mathrm{D_{\beta_1}D_{\beta_2}}\mathsf{W}_\gamma+\ldots
 + \mathsf{Q}_{\alpha}^{\beta_1\cdots\beta_n\gamma}
\mathrm{D_{\beta_1}\cdots D_{\beta_n}}\mathsf{W}_\gamma,\nonumber\\
\mathsf{\bar{W}}_{\dot{\alpha}}^{G}   &
=\mathsf{\bar{S}}_{\dot{\alpha}}  +\mathsf{\bar
{R}}_{\dot{\alpha}}^{\dot{\beta}}  \mathsf{\bar
{W}}_{\dot{\beta}}  +\mathsf{\bar{Q}}_{\dot{\alpha
}}^{\dot{\beta}_1\dot{\gamma}}  \mathrm{\bar{D}
}_{\dot{\beta}_1}\mathsf{\bar{W}}_{\dot{\gamma}}
+\mathsf{\bar{Q}}_{\dot{\alpha}
}^{\dot{\beta}_1\dot{\beta}_2\dot{\gamma}} \mathrm{\bar{D}_{\dot{\beta}_1}\bar{D}_{\dot{\beta}_2}}\mathsf{W}_{\dot{\gamma}}
+\ldots+\mathsf{\bar{Q}}_{\dot{\alpha}}^{\dot{\beta}_1\cdots\dot{\beta}_n\dot{\gamma}}
\mathrm{D_{\dot{\beta}_1}\cdots D_{\dot{\beta}_n}}\mathsf{W}_{\dot{\gamma}}. \label{wr2}%
\end{align}
Imposing supergauge invariance requires that the constitutive
supertensors $\mathsf{S}_{\alpha
}$, $\mathsf{R}_{\alpha}^{\beta}$, $\mathsf{Q}_{\alpha}^{\beta_1\gamma}$,
etc., depend only on $(x,\theta)$, the gauge superinvariant $\mathsf{X}\left(
x,\theta\right)$ that was
given by the first of Eqs. \!\!(\ref{yy}), and the superderivatives
of $\mathsf{X}\left(
x,\theta\right)$ with respect to $(x,\theta)$. In turn, the
constitutive supertensors $\mathsf{\bar{S}
}_{\dot{\alpha}}  $, $\mathsf{\bar{R}}%
_{\dot{\alpha}}^{\dot{\beta}}  $, $\mathsf{\bar
{Q}}_{\dot{\alpha}}^{\dot{\beta}_1\dot{\gamma}} $, etc., depend only on
$(x,\bar{\theta})$, the gauge superinvariant $\mathsf{Y}\left(  x,\bar{\theta}\right)$
 given by the second of Eqs. \!(\ref{yy}), and the superderivatives
 of $\mathsf{Y}\left(x,\bar{\theta}\right)$ with respect to $(x,\bar{\theta})$.
 In analogy with the case of nonlinear electromagnetism, we have $\mathsf{S}$
depending only on $(x,\theta)$ and $\mathsf{\bar{S}}$ depending
only on $(x,\bar{\theta})$, while
 $\mathsf{R}$ can depend on $\mathsf{X}\left(
x,\theta\right)$ as well as $(x,\theta)$ and $\mathsf{\bar{R}}$ can depend
on $\mathsf{Y}\left(
x,\bar{\theta}\right)$ as well as $(x,\bar{\theta})$. Finally
 the arguments of $\mathsf{Q}_{\alpha}^{\beta_1\cdots\beta_n\gamma}$ and
 $\mathsf{\bar{Q}}_{\dot{\alpha}}^{\dot{\beta}_1\cdots\dot{\beta}_n\dot{\gamma}}$
 can include up to $n$th order superderivatives of the respective supergauge invariants.

 Now, using the expansions (\ref{w2}) and (\ref{wg2}), it is straightforward
 to write down the superfield constitutive equations (\ref{wr2}) in components.

\section{Conclusion}

We have proposed a way to take an extremely general approach to nonlinear classical
electrodynamics and supersymmetric electrodynamics, for the purpose either of describing
fields and superfields in general kinds of media, or of exploring their behavior
in vacua with general properties. The framework formally takes into account
media of various types, includes non-Lagrangian as well as Lagrangian
theories, and accommodates the description of nonlocal effects. This is
accomplished through generalized constitutive equations (and constitutive tensors),
and their further generalization to include superfields. We expect future research
directions to include the detailed development of new examples within this framework.
\medskip

\section*{Acknowledgments}

S.~D. thanks Yu.~A.~Kirochkin for fruitful discussions, and Rutgers University
for hospitality during an early stage of this research.
G.~G. thanks the organizers of the meeting Quantum Theory and Symmetries 5,
Valladolid, Spain, July 2007, for the invitation to present this work. He is
also grateful for an inspiring early conversation with James Gates, about the idea
of extending nonlinear constitutive equations to superfields.


\providecommand{\newblock}{}

\end{document}